\documentclass[11pt,twoside,a4paper,cmspaper,final,collab]{cms-tdr}
\def\svnVersion{90d33ed}\def\svnDate{2026/09/27}\def\cmsCernNoTag{CERN-EP-2026-081}\def\cmsCernDate{\today}\def\cmsMessage{Published in the Journal of High Energy Physics as \href{http://dx.doi.org/10.1007/JHEP09(2026)182}{\doi{10.1007/JHEP09(2026)182}.}}
\begin{document}\cmsNoteHeader{B2G-23-003}

\newlength\cmsFigWidth
\ifthenelse{\boolean{cms@external}}{\setlength\cmsFigWidth{0.49\textwidth}}{\setlength\cmsFigWidth{0.65\textwidth}}
\newcommand{\mjj}{\ensuremath{m_\text{jj}}\xspace}
\newcommand{\mt}{\ensuremath{m_{\PQt}}\xspace}
\newcommand{\mhpm}{\ensuremath{m_{\PSHpm}}\xspace}
\newcommand{\ejets}{\ensuremath{\Pe\!\text{+jets}}\xspace}
\newcommand{\mujets}{\ensuremath{\PGm\!\text{+jets}}\xspace}
\newcommand{\PA}{\ensuremath{\mathrm{A}}\xspace}
\newcommand{\brHcs}{\ensuremath{\mathcal{B}(\PSHpm \to \PQc\PQs)}\xspace}
\newcommand{\brThb}{\ensuremath{\mathcal{B}(\PQt \to \PSHpm \PQb)}\xspace}
\newcommand{\ttX}{\ensuremath{\ttbar}+X\xspace}
\newcommand{\ttgamma}{\ensuremath{\ttbar}+\PGg\xspace}
\newcommand{\ttjets}{\ensuremath{\ttbar}+jets\xspace}
\newcommand{\wjets}{\ensuremath{\PW}+jets\xspace}
\newcommand{\dyjets}{\ensuremath{\PZ/\PGg}+jets\xspace}
\newcommand{\VV}{\ensuremath{\PW\PW,\ \PW\Z,\ \Z\Z}\xspace}
\providecommand{\DeepJet}{\textsc{DeepJet}\xspace}
\providecommand{\CP}{\ensuremath{CP}\xspace}

\ifthenelse{\boolean{cms@external}}{\providecommand{\cmsTable}[1]{#1}}{\providecommand{\cmsTable}[1]{\resizebox{\textwidth}{!}{#1}}}
\newlength\cmsTabSkip\setlength{\cmsTabSkip}{1ex}

\cmsNoteHeader{B2G-23-003}

\title{Search for light charged Higgs bosons decaying to charm and strange quarks in \texorpdfstring{\ttbar}{ttbar} events in proton-proton collisions at \texorpdfstring{$\sqrt{s}=13\TeV$}{sqrt(s)=13 TeV}}

\date{\today}

\abstract{A search is presented for a light charged Higgs boson \PSHpm in top quark pair production (\ttbar), where one of the top quarks decays to an \PSHpm and a bottom quark, while the other decays to a \PWmp boson and a bottom quark. The \PSHpm is assumed to decay into a charm and a strange quark, whereas the \PWmp boson decays into a charged lepton (electron or muon) and a neutrino. Results are reported based on proton-proton collision data at $\sqrt{s} = 13\TeV$, corresponding to an integrated luminosity of 138\fbinv.The analysis probes \PSHpm  masses in the range 40 to 160\GeV using the invariant mass spectrum of the two light jets, where light jets are defined as jets that do not satisfy the bottom quark tagging requirement. The observed yield is found to be consistent with standard model predictions. Upper limits are set on the branching fraction \brThb, with values in the range of 0.07--1.12\% (0.12--1.20\% expected) at 95\% confidence level, under the assumption that $\brHcs = 100\%$. These are the first direct limits on charged Higgs bosons produced in top quark decays for masses between 40 and 50 \GeV, and the most stringent limits to date in the 70--110\GeV range.}

\hypersetup{
pdfauthor={CMS Collaboration},
pdftitle={A search for light charged Higgs boson production in ttbar decay of proton-proton collisions at sqrt(s)= 13 TeV},
pdfsubject={CMS},
pdfkeywords={CMS, charged Higgs boson, MSSM}}
\maketitle

\section{Introduction}
\label{s:secIntro}

The discovery of the Higgs boson~\cite{Aad:2012tfa,Chatrchyan:2012ufa,ATLAS:2016neq} has opened up new opportunities to explore the scalar sector of the standard model (SM) of particle physics.
In the SM Lagrangian, the Higgs field is represented by a complex doublet under the weak isospin gauge group $SU(2)_L$.
Several beyond-the-SM theories, such as two-Higgs-doublet models (2HDMs)~\cite{Lee:1973iz,Aoki:2009ha} postulate that there can be more than a single complex doublet. Such a 2HDM introduces two Higgs doublets within $SU(2)_L$ to generate electroweak symmetry breaking.
In its unbroken form, the 2HDM framework contains eight scalar degrees of freedom. Following spontaneous symmetry breaking, three of these degrees of freedom are used to generate masses for the \PW and \PZ bosons. The remaining five appear as physical Higgs bosons: two neutral \CP-even scalars ($\PH$, $\Ph$), one neutral \CP-odd pseudoscalar ($\PA$), and two charged scalar particles ($\PSHpm$).
The branching fraction ($\mathcal{B}$) of the \PSHpm into different final states depends on the nature of the Yukawa coupling of the Higgs doublets to quarks and charged leptons.

In certain 2HDM scenarios, the charged Higgs boson can have a substantial branching fraction to a charm and a strange quark ($\PQc\PQs$),  particularly in Type II and Type X models in the low $\tan\beta$ region, and in Type Y flipped models~\cite{Aoki:2009ha} in the high $\tan\beta$ region.
Here, $\tan\beta$ denotes the ratio of the vacuum expectation values of the two Higgs doublets, $\tan\beta = v_2/v_1$, with $v_1$ and $v_2$ corresponding to the doublets that couple to down-type and up-type fermions, respectively~\cite{Aoki:2009ha}. In addition, in other 2HDMs where one Higgs doublet couples primarily to the third generation, while the second doublet couples to the first and second generations, the $\PQc\PQs$ and $\PQc\PQb$ decay channels become prominent~\cite{Altmannshofer:2016zrn}.
These channels are particularly interesting experimentally, as the final-state topology consists of two jets originating from the hadronization of quarks, allowing searches to be performed by reconstructing e.g., the invariant mass.

The upper threshold of the \PSHpm mass considered in this search is constrained by the top and bottom quark masses ($\mt, m_\PQb$), as a charged Higgs boson with mass $m_{\PSHpm} > (m_{\PQt}-m_\PQb)$ would suppress the \PSHpm production in on-shell top quark decays~\cite{LHCHiggsCrossSectionWorkingGroup:2016ypw}. Previous searches for the \PSHpm at the CERN LEP~\cite{LEPWorkingGroupforHiggsbosonsearches:2003ing,Heister:2002ev,Achard:2003gt,Abdallah:2003wd,Abbiendi:2008aa,Abbiendi:2013hk}, the Fermilab Tevatron~\cite{Abazov:2009aa,Aaltonen:2009ke}, and the CERN LHC~\cite{LHCHiggsCrossSectionWorkingGroup:2016ypw,Aad:2013hla,Aad:2014kga,Khachatryan:2015uua,CMS:2020osd} have placed constraints on its mass and branching fractions. In particular, searches at LEP excluded charged Higgs bosons with masses below 80\GeV in the context of Type II 2HDMs~\cite{Abbiendi:2013hk}. The most recent ATLAS search at $\sqrt{s}=13\TeV$, using an integrated luminosity of 140\fbinv, sets upper limits on $\brThb$ at 95\% confidence level in the range of 0.066--3.6\% for $60 < \mhpm < 168\GeV$~\cite{ATLAS:2025lightHplus}.

\begin{figure*}[htp]
\centering
\includegraphics[width=0.8\textwidth]{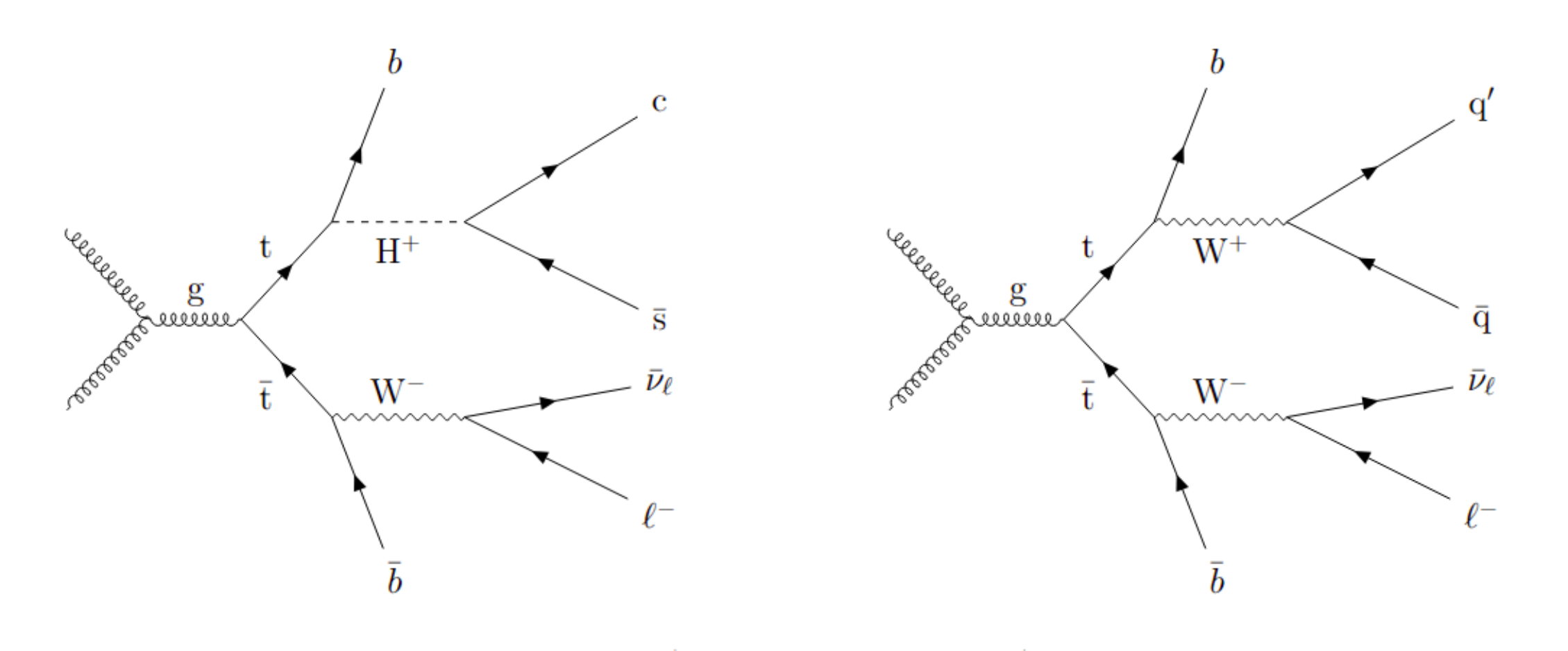}
\caption{Leading-order Feynman diagrams of \ttbar production in gluon-gluon fusion process.
   		  One possible production of \PSHpm signal from \PQt quark decay is shown in the left plot.
   		  The decay products of \PSHpm are \PQc and \PQs quarks.
   		  The diagram on the right side shows the \ttbar process in the lepton+jets channels, an irreducible SM background. }                  
\label{fig:feyn_diag_sig}
\end{figure*}

This paper presents a new search for \PSHpm in top quark decays, assuming a 100\% branching fraction to $\PQc\PQs$, in proton-proton ($\Pp\Pp$) collisions at $\sqrt{s} = 13\TeV$ with an integrated luminosity of $138~\fbinv$. The analysis strategy is similar to that described in Ref.~\cite{CMS:2020osd}, but with three main developments: (i) the use of the full 2016--2018 data set, (ii) the introduction of a multivariate categorization approach, in addition to the traditional cut-based method, to improve the signal sensitivity, as described in Section~\ref{ss:secBDT}, and (iii) an extended search range covering charged Higgs boson mass hypotheses in the 40--160\GeV\ region.
The analysis is designed to select events containing a top quark-antiquark pair, where one of the two top quarks decays to an \PSHpm and a bottom quark, with the \PSHpm further decaying to the hadronic final state. The other top quark of the \ttbar event decays to a \PWmp boson and a bottom quark, where the \PWmp boson is assumed to further decay into a lepton and the corresponding neutrino. 
The final event topology therefore comprises one charged lepton (\Pe or \PGm), missing transverse momentum attributed to one or more neutrinos, and four jets.
The dominant background arises from \ttbar production, where both top quarks decay via the SM process. Smaller contributions originate from single top quark production, multijet QCD events, and \PW and \PZ boson production in association with jets, as well as \ttX production, where $X = \PW,\,\PZ,\,\PGg,\,\PH$. A minor contribution also arises from diboson (\VV) processes.
The representative Feynman diagrams of possible \PSHpm production in  \ttbar events and the irreducible dominant SM background process are shown in Fig.~\ref{fig:feyn_diag_sig}.
    
In the following section, we describe the detector components of CMS utilized in this study.
The data and Monte Carlo (MC) samples are described in Section~\ref{s:secDataMC}, followed by a discussion of the selection criteria for the physics objects and events in Sections~\ref{s:secObjSel} and~\ref{s:secEvtSel}, respectively. Background estimation based on control samples in data is then summarized in Section~\ref{s:bkgestimn}.
We describe the signal extraction and boosted decision tree (BDT) based event categorization in Section~\ref{s:sigextract}.
The systematic uncertainties are discussed in Section~\ref{ss:secSys}. 
The results and interpretations are presented in Section~\ref{s:secResults}, followed by a summary of the paper in Section~\ref{s:secConcl}.
Tabulated results are accessible in the HEPData record for this analysis~\cite{HEPData}.

\section{The CMS detector}
\label{s:secCMS}

The CMS apparatus~\cite{CMS:2008xjf,CMS:2023gfb} is a multipurpose, nearly hermetic detector, designed to trigger on~\cite{CMS:2020cmk,CMS:2016ngn,CMS:2024psu} and identify electrons, muons, photons, and charged and neutral hadrons~\cite{CMS:2020uim,CMS:2018rym,CMS:2014pgm}.
Its central feature is a superconducting solenoid of 6\unit{m} internal diameter, providing a magnetic field of 3.8\unit{T}.
Within the solenoid volume are a silicon pixel and strip tracker, a lead tungstate crystal electromagnetic calorimeter (ECAL), and a brass and scintillator hadron calorimeter (HCAL), each composed of a barrel and two endcap sections.
Forward calorimeters extend the pseudorapidity ($\eta$) coverage provided by the barrel and endcap detectors.
Muons are reconstructed using gas-ionization detectors interleaved with the layers of the steel flux-return yoke outside the solenoid. More detailed descriptions of the CMS detector, together with a definition of the coordinate system used and the relevant kinematic variables, can be found in Refs.~\cite{CMS:2008xjf,CMS:2023gfb}.

Events of interest are selected using a two-tiered trigger system.
The first level (L1), composed of custom hardware processors, uses information from the calorimeters and muon detectors to select events at a rate of around 100\unit{kHz} within a fixed latency of 4\mus~\cite{CMS:2020cmk}.
The second level, known as the high-level trigger (HLT), consists of a farm of processors running a version of the full event reconstruction software optimized for fast processing, and reduces the event rate to a few kHz before data storage~\cite{CMS:2016ngn,CMS:2024psu}. 

\section{Data and simulation}
\label{s:secDataMC}

The analysis is based on $\Pp\Pp$ collision data collected using the CMS detector in 2016--2018, corresponding to an integrated luminosity of 138\fbinv.
Simulated signal samples are produced at next-to-leading order (NLO) following the default settings of Type II 2HDM using \MGvATNLO~v2.6.5~\cite{Alwall:2014hca}.
The \mhpm mass points are simulated from 40 to 160\GeV in 10\GeV increments, matching the resolution of the dijet invariant mass (\mjj), defined as the invariant mass of the two light jets assigned to the hadronic boson candidate.
The signal yield is derived from the inclusive \ttbar production cross section at next-to-next-to-leading order and next-to-next-to-leading logarithmic  (NNLO+NNLL) accuracy~\cite{CMS:2020osd}. The branching fraction of $\PW \to \Pell\PGn_{\Pell}$ (where $\Pell = \Pe$,\PGm,\PGt)~\cite{ParticleDataGroup:2024rcr} is applied to account for the selected final state, assuming $\brHcs = 100\%$. When the mass difference between the \PSHpm and \PW boson is smaller than one of their total widths, the interference term can be of the order of a few percent of the \PSHpm contribution, with its magnitude and sign depending on the model~\cite{MoosaviNejad:2019xxx}. This interference term is neglected in the present analysis for all values of \mhpm.
The parton-level SM \ttbar production is generated using \POWHEG~v2.0~\cite{Frixione:2007vw,Alioli:2010xd,Nason:2004rx,Alioli:2010xd}.
Three different single top quark production channels, the $s$-channel, the $t$-channel, and the $\PW$ associated production ($\PQt\PW$) are considered.
The first is generated using \MGvATNLO and the rest via \POWHEG at NLO.
The four-flavor scheme is used for the single top quark production in $s$- and $t$-channels. However, the $\PQt\PW$-channel uses the five-flavor scheme.
The production of a \PGg in association with \ttbar can be present in the \ttjets sample when the photon is misidentified as a jet. 
We follow the procedure described in Ref.~\cite{CMS:2021klw} to remove the double counting of \ttgamma events in the two MC samples.
We use the inclusive and exclusive productions of \wjets and \dyjets as part of SM background. 
These samples are generated using \MGvATNLO at NLO. 
A relatively small contribution of SM background comes from diboson production ($\PW\PW$, $\PW\PZ$ and $\PZ\PZ$) and multijet production in QCD samples enriched with muons or electrons, both generated at leading-order (LO) using \PYTHIA~8.243~\cite{Sjostrand:2014zea}.
All samples are processed with \PYTHIA~8.243 to simulate parton showers and hadronization, using the CP5 tune~\cite{CMS:2019csb}  where the matching is performed using the MLM technique~\cite{Mangano:2006rw} for the LO samples and FxFx \cite{Frederix:2012ps} for the NLO samples. All simulated samples are generated with the NNPDF3.1 NNLO parton distribution function (PDF) set~\cite{NNPDF:2017mvq}.

The CMS detector response is simulated using GEANT4~\cite{Agostinelli:2002hh}. Additional minimum bias events are simulated with \PYTHIA~8.243 and superimposed on simulated events to model the effect of additional $\Pp\Pp$ collisions within the same or adjacent bunch crossings (pileup). We use a total inelastic cross section of 69.2\unit{mb}~\cite{Sirunyan:2018nqx} to estimate the expected number of $\Pp\Pp$ interactions per bunch crossing and correct the simulation to match the corresponding distribution to that observed in data.

\section{Event reconstruction}
\label{s:secObjSel}
The primary vertex (PV) is taken to be the vertex corresponding to the hardest scattering in the event, evaluated using tracking information alone, as described in Section 9.4.1 of Ref.~\cite{CMS-TDR-15-02}.

The global event reconstruction, also called particle-flow (PF) event reconstruction~\cite{CMS:2017yfk}, aims to reconstruct and identify each individual particle in the event by optimally combining information from all CMS subdetectors. Each reconstructed particle is identified as a photon, electron, muon, charged hadron, or neutral hadron.

Muons are reconstructed by matching silicon tracker tracks to hits in the muon chambers. The reconstruction and identification efficiency exceeds 96\% across all $\eta$, with a single-muon trigger efficiency above 90\%~\cite{CMS:2018rym} at the HLT. For muons with transverse momentum (\pt) up to 100\GeV, the relative \pt resolution is about 1\% in the barrel and 3\% in the endcaps. Identification includes stringent track quality criteria with overall efficiency of 99.5\%.

The electron is detected in the tracker subdetector, whereas the energy of the electron is measured in the ECAL~\cite{CMS:2020uim}.
In the tracker, the electron track is estimated via a dedicated tracking algorithm~\cite{Adam:2005bya}, which takes into account the change of curvature of electrons due to bremsstrahlung.
Electron and photon candidates are classified using the PF algorithm with the inputs from ECAL and tracker subdetector.
An identification based on a  multivariate analysis is applied to  electrons, with a selection efficiency of 90\% for prompt electrons~\cite{CMS:2020uim}.

Lepton isolation is quantified using the relative isolation variable $I_{\text{rel}}$, defined as:
\begin{equation}
I_{\text{rel}} = \frac{1}{p_{\mathrm{T}}^\Pell} \left[ \sum_{\Delta R < 0.4(0.3)} p_{\mathrm{T}}^{\text{charged}} + \max\Bigl(0, \sum_{\Delta R < 0.4(0.3)} p_{\mathrm{T}}^{\text{neutral}} - \rho A_{\text{eff}} \Bigr) \right],
\end{equation}
where the sums run over all PF candidates within a cone of $\Delta R = \sqrt{\smash[b]{(\Delta \eta)^2 + (\Delta \phi)^2}} = 0.4$ for muons (0.3 for electrons), and the neutral component of transverse momentum due to photons and hadrons is corrected for pileup using  $\rho A_{\text{eff}}$  where $\rho$ is the diffuse offset energy density and $A_{\text{eff}}$ is the effective area.

Jets are clustered using the anti-\kt algorithm~\cite{Cacciari:2008gp,Cacciari:2011ma} as implemented in the \textsc{FastJet} package~\cite{Cacciari:2011ma}, with a distance parameter of $\Delta R = 0.4$. Contributions from pileup interactions are mitigated by discarding charged hadrons not associated with the PV and by applying corrections based on the estimated pileup energy density~\cite{CMS:2020ebo}. Jet energy corrections are derived from simulation to bring the measured response of jets to that of particle-level jets on average. In situ measurements of the momentum balance in dijet, $\text{photon}+\text{jet}$, $\PZ+\text{jet}$, and multijet events are used to correct for residual differences in the jet energy scale (JES) and the jet energy resolution (JER) between data and simulation~\cite{Khachatryan:2016kdb}. The JER amounts typically to 15--20\% at 30\GeV, 10\% at 100\GeV, and 5\% at 1\TeV. Jets are required to be isolated from selected charged leptons by $\Delta R > 0.4$. The missing transverse momentum vector \ptvecmiss is computed as the negative vector sum of the transverse momenta of all PF candidates, and its magnitude is denoted as \ptmiss. It includes corrections to the JES~\cite{CMS:2019ctu}.

Flavor tagging is performed using the \DeepJet algorithm~\cite{Bols:2020bkb}, which employs a deep neural network to identify jets originating from \PQb, \PQc, or light-flavor quarks or gluons. The discriminator distributions in simulation are calibrated to data using scale factors derived from control samples in data~\cite{CMS:2021scf}. At the medium working point, the identification efficiency for \PQb jets is approximately 80\%, with a light-jet mistag rate below 1\%.

\section{Event selection}
\label{s:secEvtSel}

Each selected event contains exactly one isolated lepton (\Pe or \PGm) that satisfies the single-electron or single-muon HLT and the corresponding offline requirements. For muons, the trigger requires $\pt > 24\GeV$ for 2016 and 2018, and $\pt > 27\GeV$ for 2017. After offline reconstruction, the selected muons must satisfy $\abs{\eta} < 2.4$, $I_\text{rel}^\PGm < 0.15$, and the vertex impact parameters $\abs{d_{z}} < 0.10\cm$ and $\abs{d_{xy}} < 0.02\cm$, where $d_z$ and $d_{xy}$ denote the longitudinal and transverse components, respectively, of the distance of closest approach of the track to the PV. An offline \pt threshold of 26\GeV is applied, except for the 2017 data-taking, where 30\GeV is used, with an identification efficiency of 99.5\%~\cite{CMS-DP-2020-040}. The $I_\text{rel}^\PGm < 0.15$ selection has an efficiency close to 95\% for muons with \pt $>27\GeV$.

For electrons, the trigger requires $\pt > 27\GeV$ in 2016 and $\pt > 32\GeV$ in 2017 and 2018. After offline reconstruction, the selected electrons must satisfy $\abs{\eta} < 2.4$, excluding the ECAL transition region ($1.44 < \abs{\eta} < 1.57$), and pass isolation requirements of $I_\text{rel}^\Pe < 0.08$ (barrel) or $<$0.07 (endcap). Prompt electrons must satisfy $\abs{d_{z}} < 0.10\,(0.20)$\,cm and $\abs{d_{xy}} < 0.05\,(0.10)$\cm for the barrel (endcap) regions. Offline \pt thresholds are 30\GeV (2016) and 35\GeV (2017--2018), motivated by the change in thresholds of the trigger selection.

To reduce contamination from \PZ boson decays and semileptonic heavy-flavor decays, events are vetoed if a second lepton (muon or electron) is found with $\pt > 15\GeV$, $\abs{\eta} < 2.4$, and $I_\text{rel}^\PGm < 0.4$ or $I_\text{rel}^\Pe < 0.3$. These veto selection thresholds are less stringent than those applied to the primary (candidate) muon or electron, to suppress background while preserving signal efficiency.

Events must contain at least four jets with $\pt > 25\GeV$ and $\abs{\eta} < 2.4$. The missing transverse momentum is required to satisfy $\ptmiss > 20\GeV$. Events with spurious high \ptmiss due to reconstruction failures, detector malfunctions or non-collision backgrounds are removed by dedicated selection conditions~\cite{CMS:2019ctu}. Flavor tagging criteria are applied to identify signal-like topologies as in Fig.~\ref{fig:feyn_diag_sig}. Events must contain at least two \PQb-tagged jets passing the medium working point of the \DeepJet algorithm. Additional selections for categorization, including a use of charm quark tagging (\PQc-tagging) discriminators, are discussed in Section~\ref{ss:secBDT}.

\section{Background estimation}
\label{s:bkgestimn}
The dominant background after applying the event selection arises from SM \ttbar and \ttX processes, contributing approximately 88--92\% of the total background depending on the three charm tagging categories, two lepton channels and three data-taking years. 
Other simulated background sources due to single top quark production, $\PW$+jets, $\PZ/\PGg$+jets, and diboson processes together account for about 4--6\% of the total background.  
All of these backgrounds are estimated from simulation and subsequently constrained in a binned likelihood fit to data, where their normalizations and shapes are allowed to vary within their associated systematic uncertainties, whereas the QCD multijet background, which accounts for the remaining background contribution, is estimated from control regions in data and its magnitude is determined within the same fit, as it is not modeled with sufficient accuracy in simulation.

The signal region (SR) is defined by the set of events selection criteria described in Section~\ref{s:secEvtSel}.  
The control region (CR) is defined similarly, except the lepton is required to fail the isolation requirement.  
This region is enriched in QCD multijet events and is used to estimate the QCD background contribution from data.
This contribution is estimated by an AB method (a reduced version of the ABCD procedure~\cite{CMS:2024onh}), where regions A and B are the SR and CR, respectively. The QCD multijet contribution is first estimated in the CR by subtracting the yield of the simulated non-QCD backgrounds from the data.
A simultaneous binned likelihood fit is performed to determine the QCD background rate and extract its contribution in the SR.  
The yield in the CR is scaled by a dedicated rate parameter (described in Section 4.3 of \cite{CMS:2024onh}), which is determined from the fit and extrapolates the QCD contribution to the SR. 
To validate this method, the shapes of the \mjj distributions in QCD multijet events in the SR and CR regions are compared using QCD-enriched MC samples, and good agreement is observed, supporting the assumption that the CR provides a reliable template for the SR shape.  
A variation of 50\% of the  $I_\text{rel}$ boundaries is applied to estimate the  upper or lower bound of the systematic uncertainty in the background estimation.

\section{Signal extraction}
\label{s:sigextract}
\subsection{Invariant mass distribution}
\label{s:kinfit}
The signal sensitivity in the final state is enhanced through a kinematic fit to the variables characterizing the event topology ~\cite{BaBar:2003org,BaBar:2004tuo,DHondt:2006iej}. This procedure reduces ambiguities in the reconstruction of intermediate resonances, particularly by mitigating the high-mass tails often observed in the \mjj spectrum prior to the fit.
The high-mass tails in the \mjj spectrum arise because of  ambiguities in the correct association of \PQb jets in events where there are more than two such jets.
Such tails are suppressed after the application of kinematic fitting.

The kinematic fit is based on a least-squares minimization approach, implemented on an event-by-event basis using Lagrange multipliers. These multipliers enforce physical constraints dictated by the decay topology. Specifically, the four-momentum $(\pt, \eta, \phi, m)$ of the lepton and jets, along with the \ptmiss, are used as inputs together with their associated detector resolutions.
For leptons, m is taken to be the rest mass, and for jets, the mass is calculated according to the procedure described in~\cite{CMS:2017wyc}.
Additionally, \PQb tagging discriminators are employed to identify the \PQb-tagged jets used in the fit. The jets which are not \PQb-tagged are termed as light jets.

To address ambiguities in \PQb jet assignment, all possible combinations of the two \PQb jets are considered. The algorithm evaluates which \PQb jet is associated with the lepton--neutrino system and which is associated with the two light jets. A correct association of \PQb jets is important to reduce the tail contribution in the invariant mass distribution of light jets due to the SM processes other than desired \ttbar semileptonic decay. The configuration is chosen by selecting the combination that best matches the expected mass of the \ttbar decay chain~\cite{BaBar:2003org,BaBar:2004tuo,DHondt:2006iej}.

The selection is based on three primary invariant mass criteria:
\begin{enumerate}
    \item The invariant mass of the lepton and neutrino (\ptmiss) system is compared to the nominal \PW~boson mass.
    \item The invariant mass of the lepton, neutrino, and \PQb-tagged jet system is evaluated against the expected top quark mass.
    \item The invariant mass of the two light jets and the remaining \PQb-tagged jet is evaluated against the expected top quark mass.
\end{enumerate}
During the minimization, the kinematic parameters (\ie, $\pt$, $\eta$, $\phi$) of the reconstructed particles are varied within their respective resolutions. The neutrino transverse momentum is taken from \ptmiss, while its longitudinal component is treated as a free parameter and determined from the fit. The masses of the parent particles (W boson and top quark) are also allowed to vary within their natural widths under a Gaussian constraint.

The goodness of the fit is quantified by a $\chi^2$ variable, constructed from the deviations between the adjusted and measured kinematic quantities, divided by their resolutions.

If more than two \PQb-tagged or light jets are present, the fit is repeated for all relevant jet--to--parton assignments, and the corresponding $\chi^2$ values are stored. Among all jet combinations, the one with the lowest $\chi^2$ is selected. The two selected light jets are then interpreted as the decay products of the \PSHpm), and their invariant mass is taken as the reconstructed mass of the \PSHpm system.
The efficiencies of the kinematic fit selection are 83\%, 80\%,86\% and 78\% for the SM \ttbar, \PSHpm(50\GeV), \PSHpm(100\GeV) and \PSHpm(150\GeV), respectively.

\subsection{Event categorization}
\label{ss:secBDT}

The event categorization strategy is designed to enhance the signal sensitivity by optimally exploiting the charm-tagging information and multivariate correlations among kinematic and topological variables. The analysis is performed separately in the \mujets and \ejets final states. Two complementary approaches are employed in each channel: the main multivariate categorization using a BDT, and an alternative cut-based categorization referred to as the cut-based method.

The cut-based method provides an independent cross-check of the results obtained with the BDT-based categorization. In this method, the two light jets from the jet assignment corresponding to the lowest $\chi^{2}$ value from the kinematic fit are classified according to their charm tagging properties~\cite{CMS:2021scf}. Charm tagging is performed using dedicated \DeepJet algorithms that classify jets based on their likelihood of originating from a charm quark. These algorithms provide two main discriminators: CvsL, which separates charm jets from light (\PQu, \PQd, \PQs, \Pg) jets, and CvsB, which separates charm jets from bottom jets.

Three categories are defined based on efficiency and misidentification rates~\cite{CMS:2021scf}: loose, medium, and tight. The categorization is based on the highest charm tagging score among the two light jets, obtained from a simultaneous selection cut on the CvsL and CvsB discriminators:
\begin{itemize}
    \item {Loose}: passes the loose working point but fails the medium one.
    \item {Medium}: passes the medium working point but fails the tight one.
    \item {Tight}: passes the tight working point.
\end{itemize}

The loose category provides higher efficiency at the cost of increased misidentification, whereas the tight category has the lowest misidentification rate, with reduced efficiency.

Events in which neither of the two light jets pass any charm tagging working point are rejected. The remaining events are exclusively categorized based on the tightest charm tagging requirement satisfied by the pair of jets.

\begin{table*}[htb]
\centering
\topcaption{The input variables used for event categorization in the BDT}
\label{Variables}
\renewcommand{\arraystretch}{1.2}
\cmsTable{
\begin{tabular}{ll}
\hline 
{Category of variable} & {Description} \\
\hline
\multirow{7}{*}{Kinematic}
& \pt of the leading light jet. \\
& \pt of the subleading light jet. \\
& Scalar sum of the \pt of the four jets. \\
& Invariant mass of the two \PQb-tagged jets. \\
& Invariant mass of the reconstructed \ttbar system. \\
& \pt of the \PQb-tagged jet originating from hadronic top quark decay. \\
& \pt of the \PQb-tagged jet originating from semileptonic top quark decay. \\[\cmsTabSkip]
\multirow{4}{*}{Angular}
& $\Delta\phi$ between the two light jets. \\
& $\Delta\phi$ between the leading light jet and the hadronic \PQb-tagged jet. \\
& $\Delta\phi$ between the lepton and the \PQb-tagged jet from semileptonic top decay. \\
& $\Delta R$ between the leading light jet and the lepton side \PQb-tagged jet. \\[\cmsTabSkip]
\multirow{6}{*}{c tagging}
& CvsL score of the leading light jet. \\
& CvsL score of the subleading light jet. \\
& CvsL score of the \PQb-tagged jet from hadronic decaying top. \\
& CvsL score of the \PQb-tagged jet from semileptonic decaying top. \\
& CvsB score of the leading light jet. \\
& CvsB score of the \PQb-tagged jet from semileptonic decaying top. \\[\cmsTabSkip]
$\cos \theta^{*}$
& $\cos \theta^{*}$ between the hadronic top quark and the light jet with higher CvsL score in the rest frame of the \PSHpm candidate. \\
\hline
\end{tabular}
}
\end{table*}

\begin{figure*}[htb]
\centering

\includegraphics[width=0.45\linewidth]{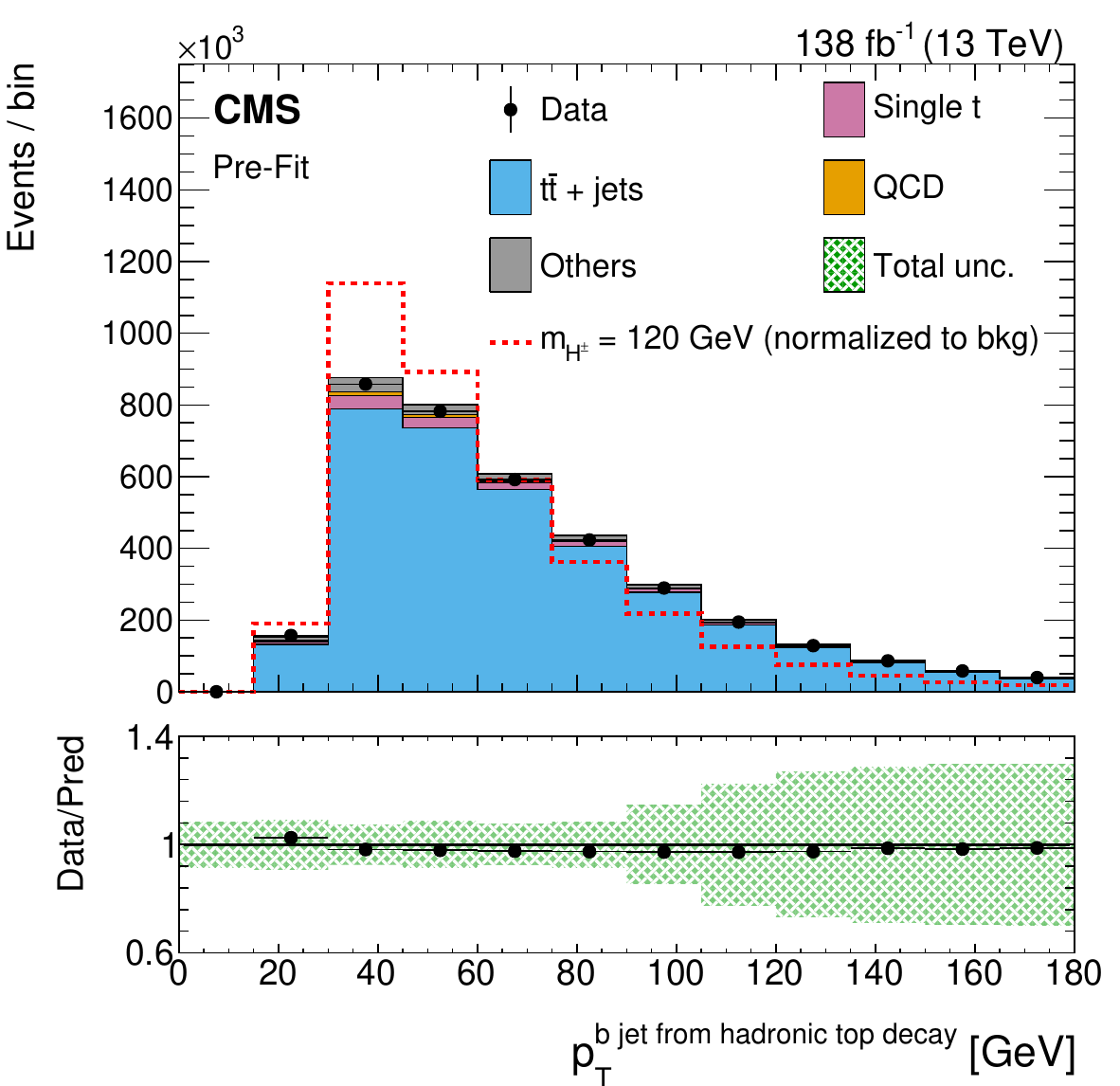}
\includegraphics[width=0.45\linewidth]{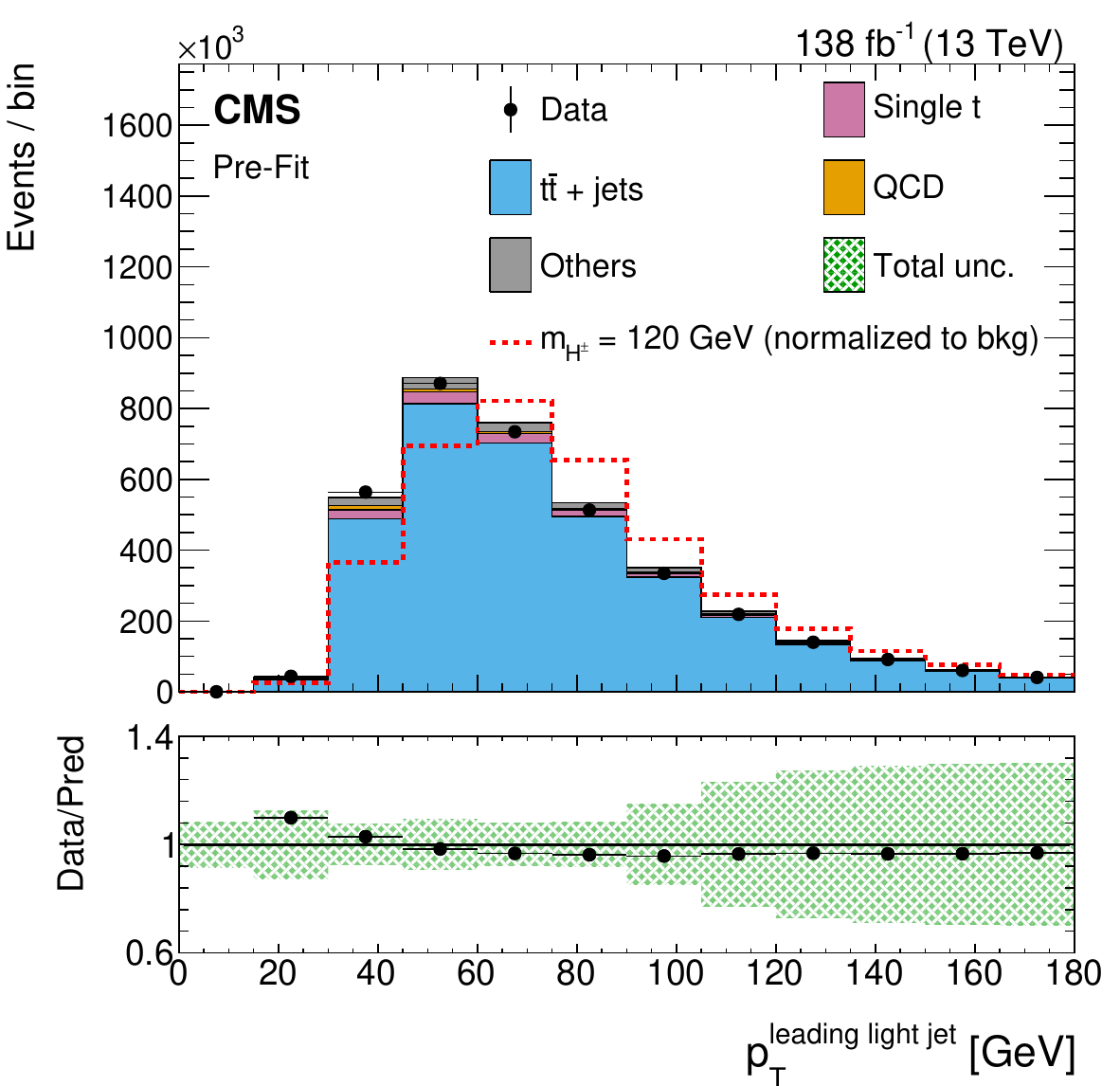}

\includegraphics[width=0.45\linewidth]{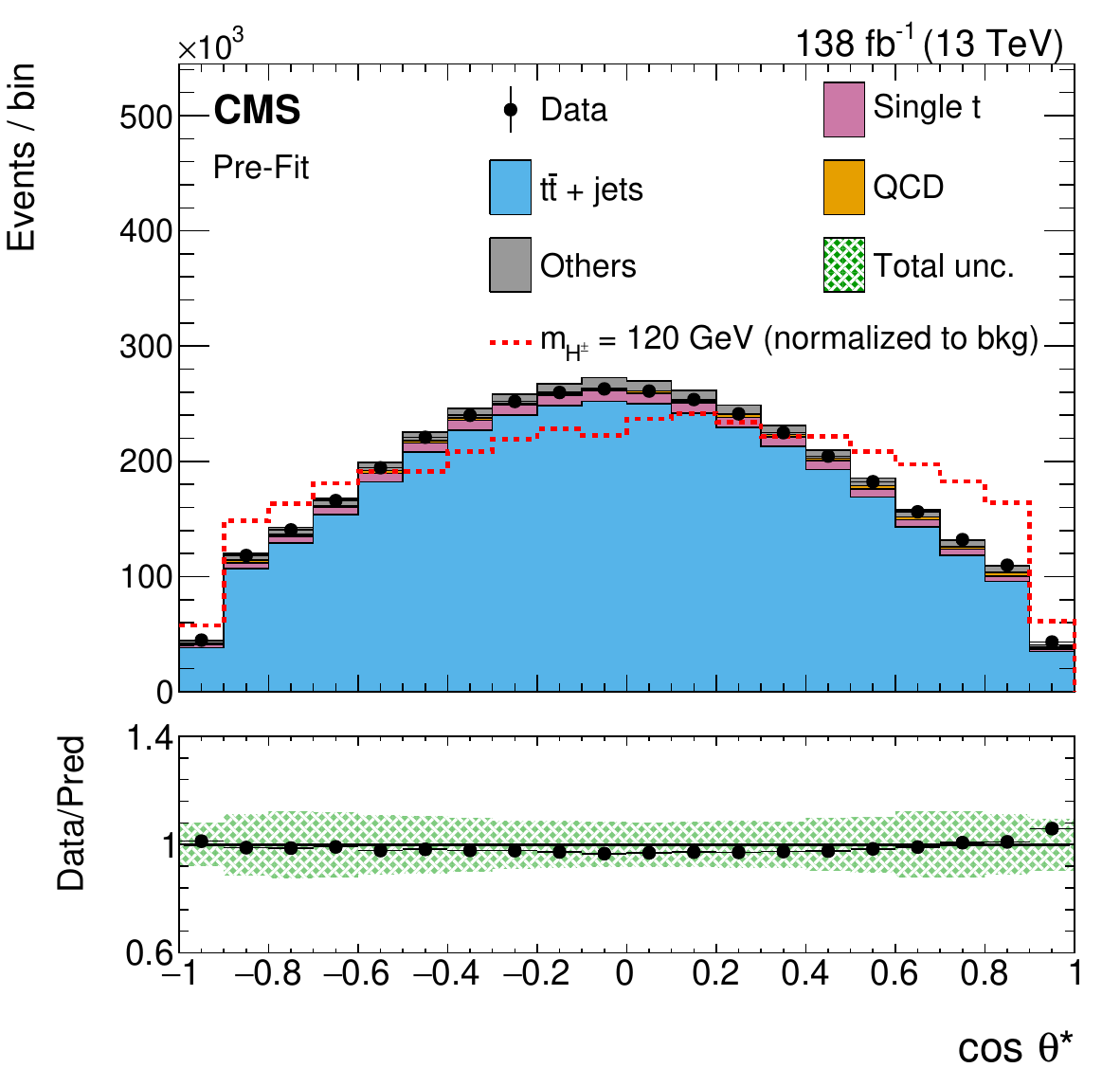}
\includegraphics[width=0.45\linewidth]{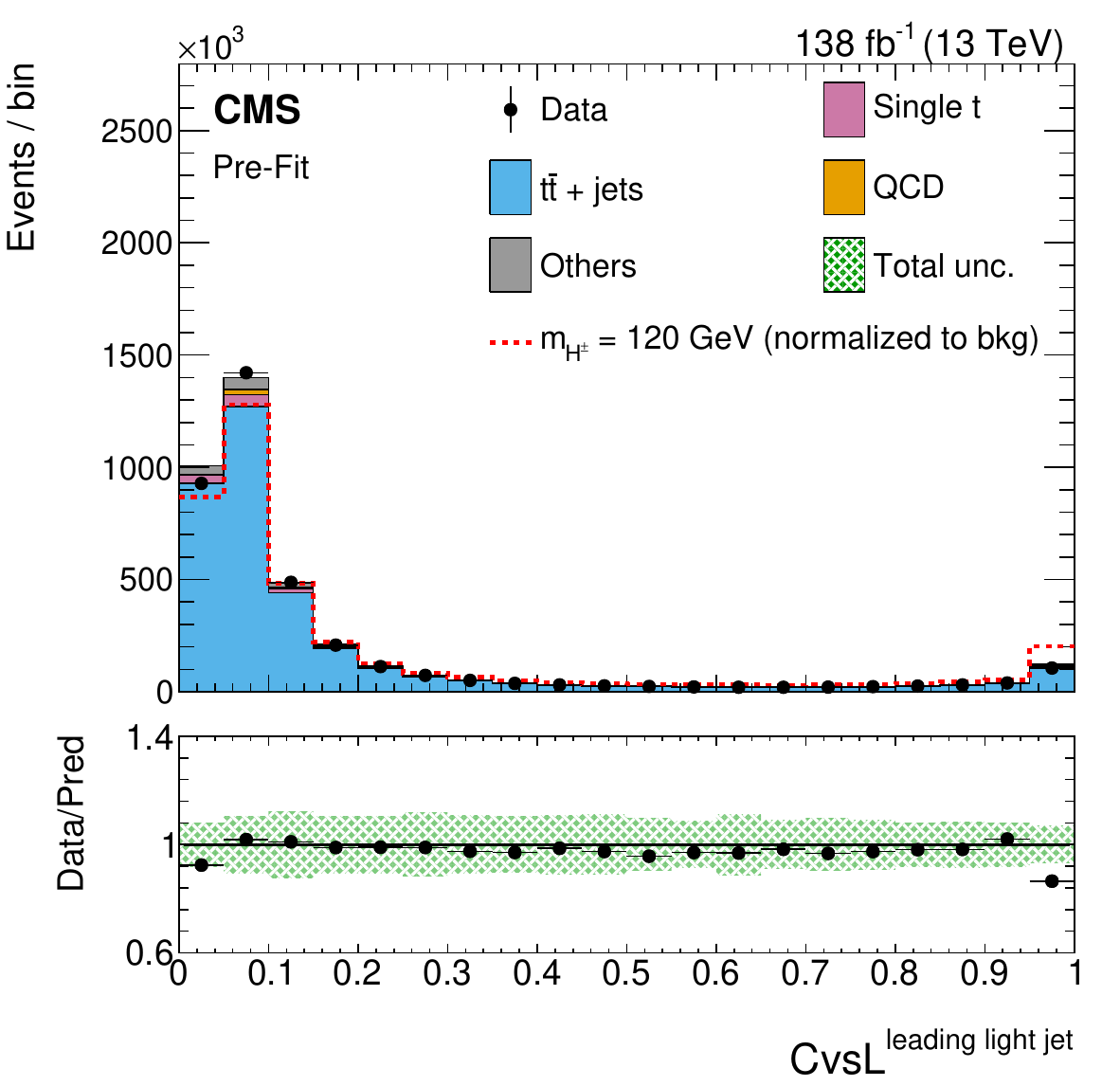}

\caption{
Distributions of selected input variables used in the BDT training for the combined data set.
Shown are the pre-fit distributions of the \pt of the \PQb-tagged jet from the hadronic top quark decay (upper left),
the \pt of the leading light jet (upper right),
the $\cos\theta^\ast$ discriminant variable (lower left),
and the CvsL discriminator value of the leading light jet (lower right). The lower panels display the ratio of data to the predicted background, with the shaded band representing the total uncertainty in the prediction. The potential signal distribution is normalized to the background yield.
}
\label{fig:bdt_input_vars}
\end{figure*}

In the multivariate method, a BDT is used to improve the separation between signal and background events. This helps to refine the categorization by effectively utilizing the available information to enhance the signal-to-background ratio in the final event samples. Since the SM \ttbar events comprise close to 90\% of the total background in the SR, the BDT is trained using only SM \ttbar as background and \PSHpm as signal. Eighteen variables that discriminate between the \PSHpm and \ttbar processes are used as inputs to the BDT. These variables are listed in Table~\ref{Variables}, and distributions of a representative subset of these input variables are shown in Fig.~\ref{fig:bdt_input_vars}. One of the input variables is $\cos\theta^{*}$, defined as the cosine of the angle between the hadronic top quark and the light jet with the higher CvsL score, evaluated in the rest frame of the \PSHpm candidate.

 Two separate BDT trainings are performed for the low- (40--70\GeV) and high-mass (80--160\GeV) points.
 The dijet invariant mass of the two light jets (the \PQc and \PQs candidates) is used as the final discriminator. Therefore, it is not included as an input variable to BDT, which is used for event categorization. However, the BDT may still have a dependence on the dijet invariant mass because of a possible correlation of the dijet invariant mass with other BDT input variables. In order to remove any dependence of the BDT on \mjj, the background samples in training are weighted so that they match the \mjj distributions of the combined signal samples. All signal mass points (i.e, 40--70 \GeV and 80--160 \GeV) are combined for the low- and high-mass sets of training, respectively.  The training is performed on the approximately equal sample of signal and SM \ttbar events ($\sim$500\,000 events each). The weighting procedure also normalizes the weights of signal and background events so that their sum is equal, in order to avoid any potential biases due to training on unbalanced samples. The input events are randomly assigned to either the training sample or the test sample, the latter being used as a validation set to check for over training. The BDT is trained using the gradient boosting algorithm of the \textsc{scikit-learn} package~\cite{Pedregosa:2012toh}. The hyper-parameters used in the training process include a learning rate of 0.1, a number of estimators equal to 800, a subsample fraction of 0.6, and a maximum tree depth of 4 for the high-mass model; 600 estimators and a depth of 3 for the corresponding low-mass model. These hyper-parameters were optimized through a systematic search over a range of values to maximize the model performance on the validation set, using cross-validation to avoid overfitting~\cite{Kohavi:1995cv}.

\begin{figure}[hbt!]
  \centering
  \includegraphics[width=0.48\textwidth , height=6cm]{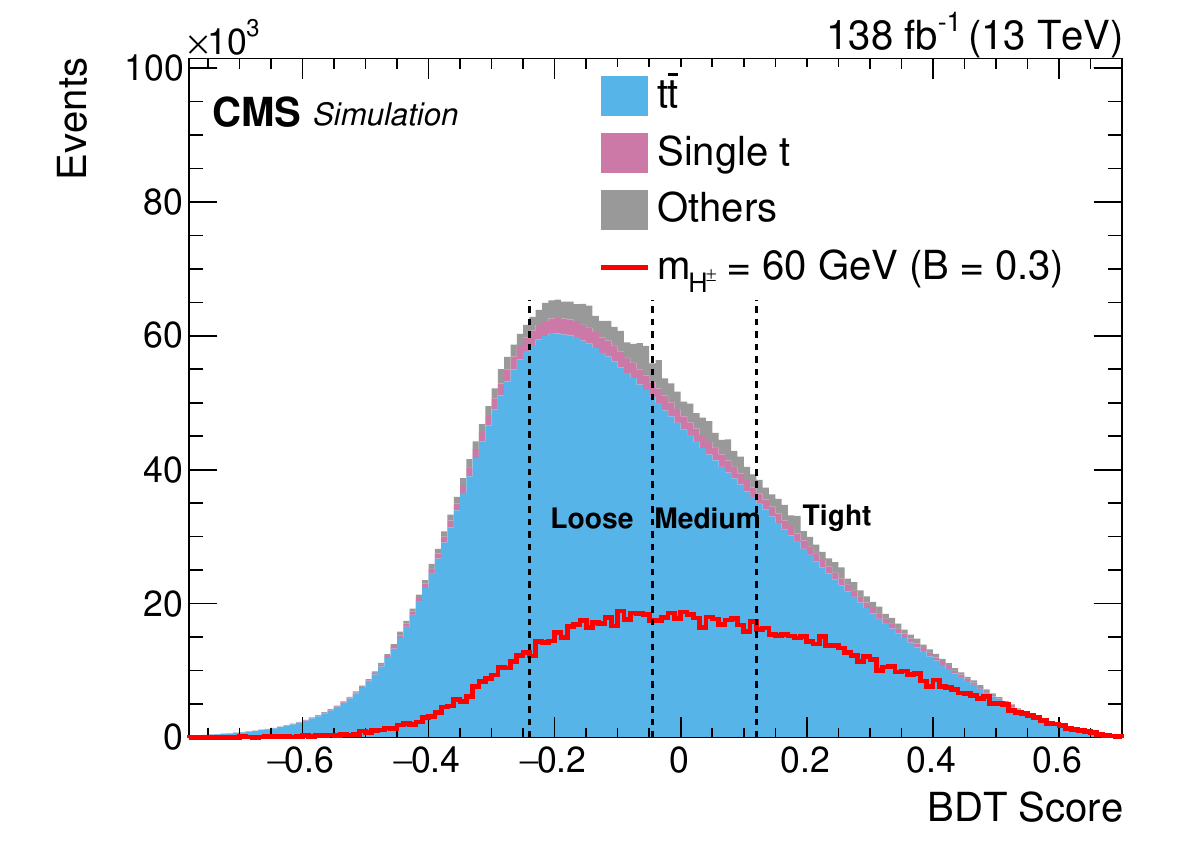}
  \hfill
  \includegraphics[width=0.48\textwidth , height=6cm]{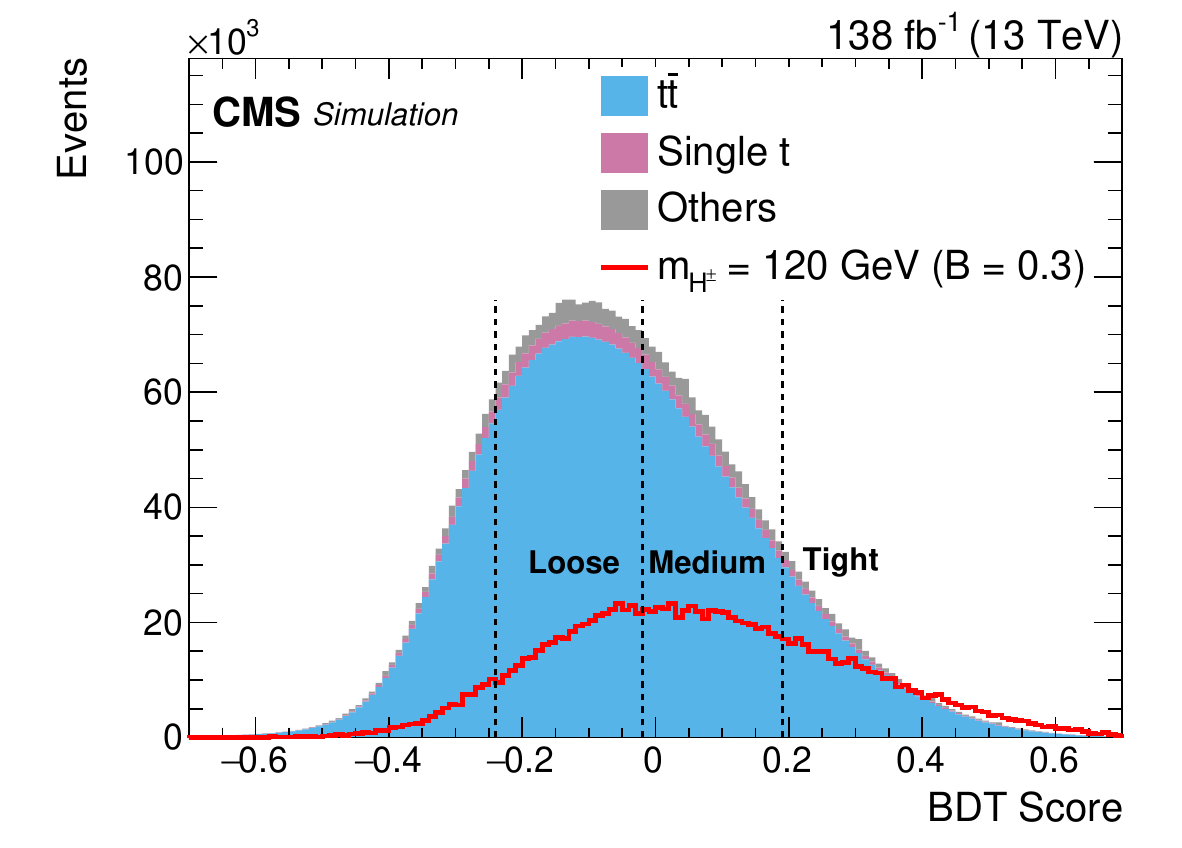}
  \caption{BDT output score distribution for the backgrounds and signal processes for the (left) low-mass and (right) high-mass models. The vertical lines correspond to the working points used for categorization. The signal distribution is normalized to $\brThb = 30\%$ for visualization purposes.}
  \label{categories}
\end{figure}

The BDT was validated by comparing the distributions of input variables and the output score between the training and validation data sets to ensure consistency and detect potential over-fitting. Additionally, data-to-simulation agreement was verified for the input variables and the BDT output score in a CR defined with inverted \ptmiss requirements.
To increase the sensitivity of the analysis, the events in the SR are categorized based on the BDT output distribution into three exclusive categories: loose, medium, and tight. Events that do not satisfy any of these category requirements are discarded. 
The working points for categorization are chosen to maximise the separation in the signal-to-background ratio between the categories, while also retaining the same number of signal events in each category. Figure~\ref{categories} shows the distribution and working points of the BDT output used for the categorization.

\section{Systematic uncertainties}
\label{ss:secSys}

A study has been conducted to estimate the systematic uncertainties due to various sources, such as those originating from detector calibration, reconstruction procedure, and uncertainties of theory models.
We quote systematic uncertainties corresponding to the variations in the binned \mjj distributions used in the fit, obtained by varying each source of systematic uncertainty with respect to the nominal prediction.
For systematic uncertainty categories with multiple independent sources, such as the JES and JER~\cite{Khachatryan:2016kdb}, the quoted values represent the maximum average variation in the \mjj shape from the single most dominant sub-source. The SM \ttbar distribution is used throughout to evaluate these uncertainties, unless stated otherwise.

A summary of the various uncertainty sources and their impact on the \mjj distribution is provided in Table~\ref{tab:systematics_summary}.

\begin{table*}[htb]
\centering
\caption{Summary of the main sources of systematic uncertainties and their typical impact on the \mjj distribution. The ranges reflect variations across different data-taking years and event categories.}
\label{tab:systematics_summary}

\begin{tabular}{l c}
\hline
Source of uncertainty & Typical impact [\%] \\
\hline
Pileup reweighting & 2--3 \\
JES and JER & 4--5 \\
\PQb and \PQc tagging  & 3--17 \\
Renormalization and factorization scale & 3--11 \\
ISR and FSR modeling & 1--2 \\
Underlying event tune & 4 \\
Top quark mass & 4--5 \\
Cross section normalization & 4--7 \\
\hline
\end{tabular}
\end{table*}

The integrated luminosities for the 2016, 2017, and 2018 data-taking years have 1.2--2.5\% individual uncertainties~\cite{CMS-LUM-17-003,CMS-PAS-LUM-17-004,CMS-PAS-LUM-18-002}, while the overall uncertainty for the 2016--2018 period is 1.6\%.

The central value of the pp total inelastic cross section is varied within its uncertainty of 4.6\% to obtain the corresponding variation in pileup reweighting.

During the 2016 and 2017 data-taking, a gradual shift in the timing of the inputs of the ECAL L1 trigger in the region at $\abs{\eta} > 2.0$ caused a specific trigger inefficiency. For events containing an electron (a jet) with \pt larger than $\approx$50\GeV ($\approx$100\GeV), in the region $2.5 < \abs{\eta} < 3.0$ the efficiency loss is $\approx$10--20\%, depending on \pt, $\eta$, and the data-taking period.
Correction factors were computed from data and applied to the acceptance evaluated by simulation. This uncertainty is found to have an impact of less than 1\% to the \mjj distribution. 

The correction factors for the identification, isolation, and trigger efficiencies for muons and electrons are obtained using the tag-and-probe method~\cite{CMS:2020uim}. The combined effects of systematic uncertainties due to lepton correction factors is found to be $1\%$ for both muons and electrons.

Uncertainties from the JES and JER are assessed by constructing alternative templates with varied jet momenta. JES variations are applied by shifting the jet \pt up and down as a function of \pt and $\eta$, while JER uncertainties are modeled by varying the smearing applied in simulation. These variations are propagated to all derived observables, including \ptmiss and invariant masses. The resulting templates are used to evaluate the impact of JES/JER uncertainties.

Various sources of systematic uncertainties associated with the \PQb and \PQc tagging calibrations are evaluated.
The uncertainties that are correlated with other sources, such as pileup, jet energy corrections, and \ptmiss, as well as renormalization and factorization scales, are treated consistently within the corresponding systematic uncertainty evaluation~\cite{CMS:2021scf}. There are three sets of uncertainties due to the \PQb and \PQc tagging shape calibration that are not correlated with other sources. The dominant contribution arises from the limited size of the simulated samples used in the \PQc tagging calibration. The selection of \PQc jets is also sensitive to the flavor composition of background jets, such as \PQb and light jets, and this effect is taken into account as a separate source of uncertainty. Further uncertainties arise from the interpolation and extrapolation procedures applied in the \PQc tagging calibration. These uncertainties are treated as uncorrelated across event categories and data-taking periods.

The uncertainties associated with the modeling of simulated processes arise from missing higher-order corrections in theoretical predictions. These uncertainties are estimated by independently varying the renormalization and factorization scales by factors of 2 and 0.5. An envelope is constructed from the resulting variations, excluding configurations where one scale is increased while the other is decreased (\ie, one varied by 2 and the other by 0.5). The resulting uncertainties are treated as fully correlated across all data-taking periods. An additional uncertainty associated with the choice of PDFs is evaluated using the standard deviation across a set of MC replicas, following the prescription detailed in Ref.~\cite{Butterworth:2015oua}. This approach accounts for variations in the predicted distributions arising from the limited knowledge of the proton structure, and is applied to both signal and background samples where relevant.

Uncertainties associated with the modeling of parton shower scales for initial- and final-state radiation (ISR and FSR) are evaluated using dedicated MC samples. For ISR, the parton shower scale in \PYTHIA~8.243 is varied by scaling the default value up by a factor of 2 and down by 0.5. For FSR, the scale is modified by factors of $\sqrt{2}$ and $1/\sqrt{2}$, respectively, following the tuning procedure described in Ref.~\cite{Skands:2014pea}. Parameters of the CP5 underlying event tune are varied to obtain associated uncertainties.
An uncertainty of 4\% is assigned to cover observed differences between the results obtained with the PYTHIA and HERWIG parton shower models~\cite{CMS:2024irj,CMS-DP-2020-019}.

The systematic uncertainty due to the top quark mass is evaluated by varying the nominal value of $m_{\PQt} = 172.5\GeV$ by $\pm1.0\GeV$. The matching between the matrix element calculation in the \textsc{Powheg} generator and the \PYTHIA parton shower is governed by a parameter that controls the damping of real emissions in the matrix element ($h_{\text{damp}}$). Variations of this parameter are used to estimate the corresponding modeling uncertainty~\cite{CMS-PAS-TOP-16-021}.

Systematic uncertainties related to the normalization of simulated samples arise from the theoretical uncertainties in the production cross sections, as reported in Ref.~\cite{CMS:2020osd}. The statistical uncertainty arising from the finite size of MC samples is incorporated using the Barlow--Beeston ``lite'' method~\cite{Barlow:1993dm}, which treats the per-bin MC uncertainty as a nuisance parameter in the fit. This approach ensures a proper treatment of bin-by-bin fluctuations, particularly in regions with low event yields or high \pt. The systematic uncertainties associated with the QCD multijet background are described in Section~\ref{s:bkgestimn}.
A summary of the relative impact of the different groups of systematic uncertainties on the uncertainty in the extracted $\mathcal{B}_{\PSHpm}$ for two different mass points is provided in Table~\ref{tab:uncertainty_breakdown_custom}.

\begin{table*}[htb]
\centering
\caption{Breakdown of systematic and statistical uncertainties in the extracted $\mathcal{B}_{\PSHpm}$ for two signal mass points. Each percentage represents the contribution of the corresponding nuisance parameter group relative to the total uncertainty when all sources are combined in quadrature.}
\label{tab:uncertainty_breakdown_custom}

\begin{tabular}{l c c}
\hline
Category & \multicolumn{2}{c}{$\mathcal{B}(\PSHpm)$ [\%]} \\
         & $\PSHpm_{80}$ & $\PSHpm_{150}$ \\
\hline
JES and JER        & 58 & 39 \\
\PQb and \PQc jet tagging      & 48 & 38 \\
Modeling                & 29 & 35 \\
Int. luminosity               &  8 & 11 \\
Normalization            &  6 & 10 \\
Statistical              & 12 & 39 \\
\hline
\end{tabular}

\end{table*}

\section{Results}
\label{s:secResults}

\begin{figure}[htb]
    \centering
    \includegraphics[width=0.49\textwidth]{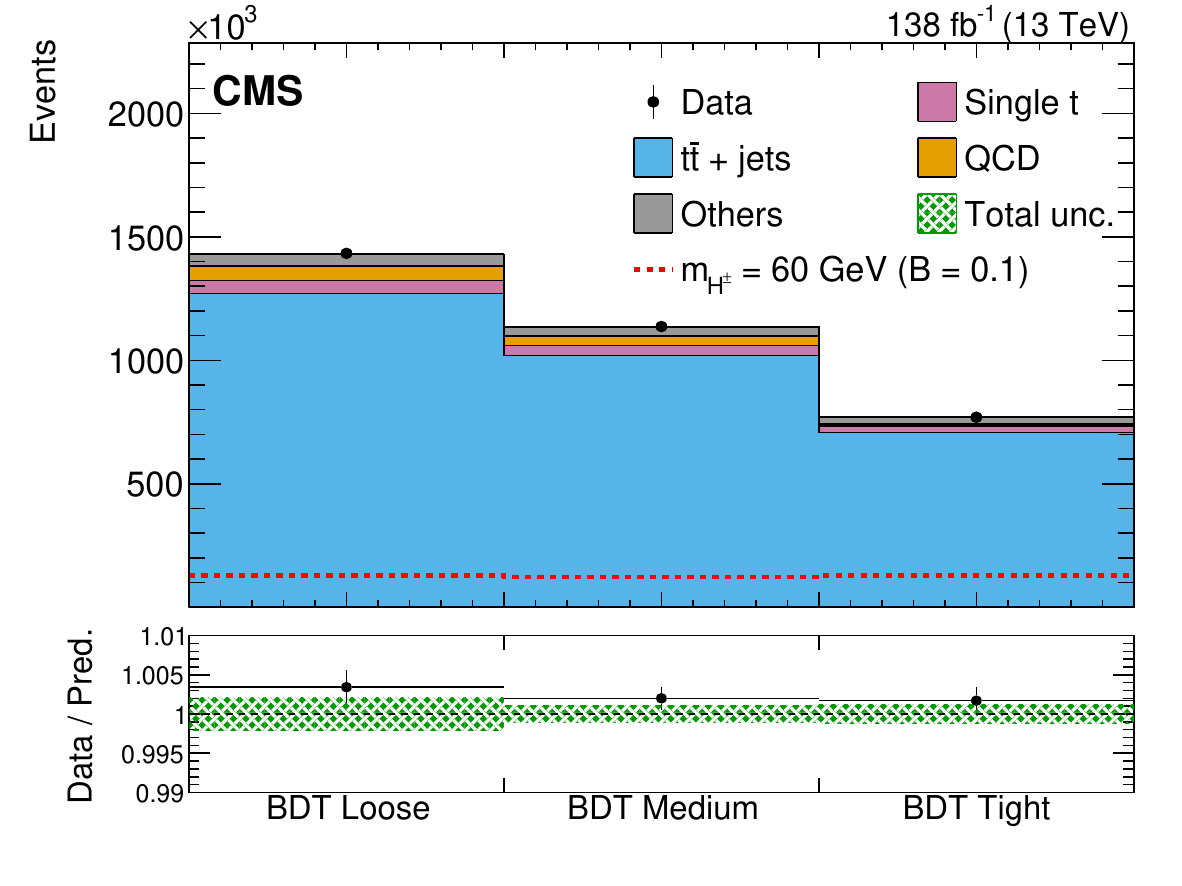}
    \includegraphics[width=0.49\textwidth]{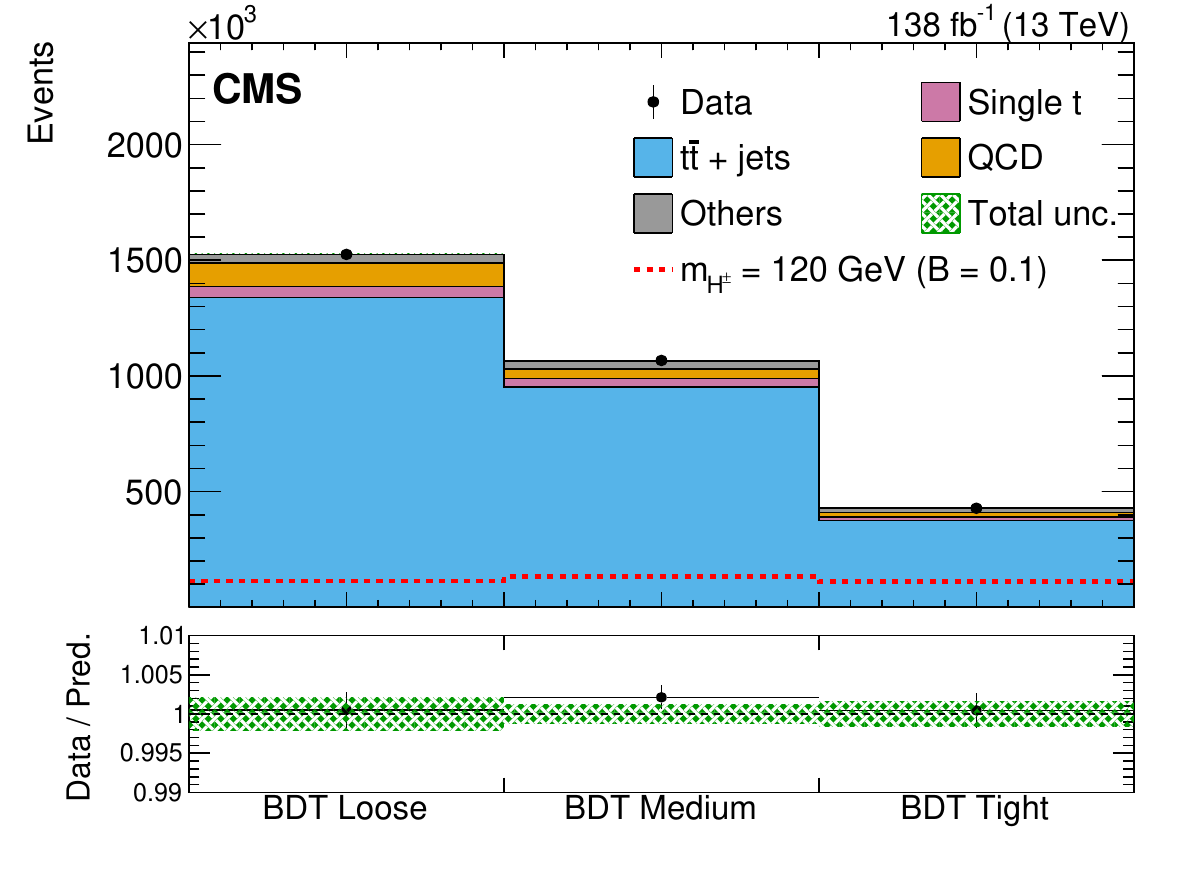}
    \caption{
        Post-fit event yields of the BDT categories after the background-only fit.
        The lower panel shows the ratio of data to the predicted background, with the shaded band indicating the total uncertainty in the prediction, including both statistical and systematic contributions. For comparison, the potential signal yields with $\brThb=10\%$ are shown for charged Higgs boson masses of $\mhpm=60\GeV$ (left) and $\mhpm=120\GeV$ (right) corresponding to the low-mass and high-mass models, respectively.}
    \label{fig:eventYield}
\end{figure}

\begin{figure*}
\centering
\makebox[\textwidth][c]{
  \includegraphics[width=0.48\linewidth]{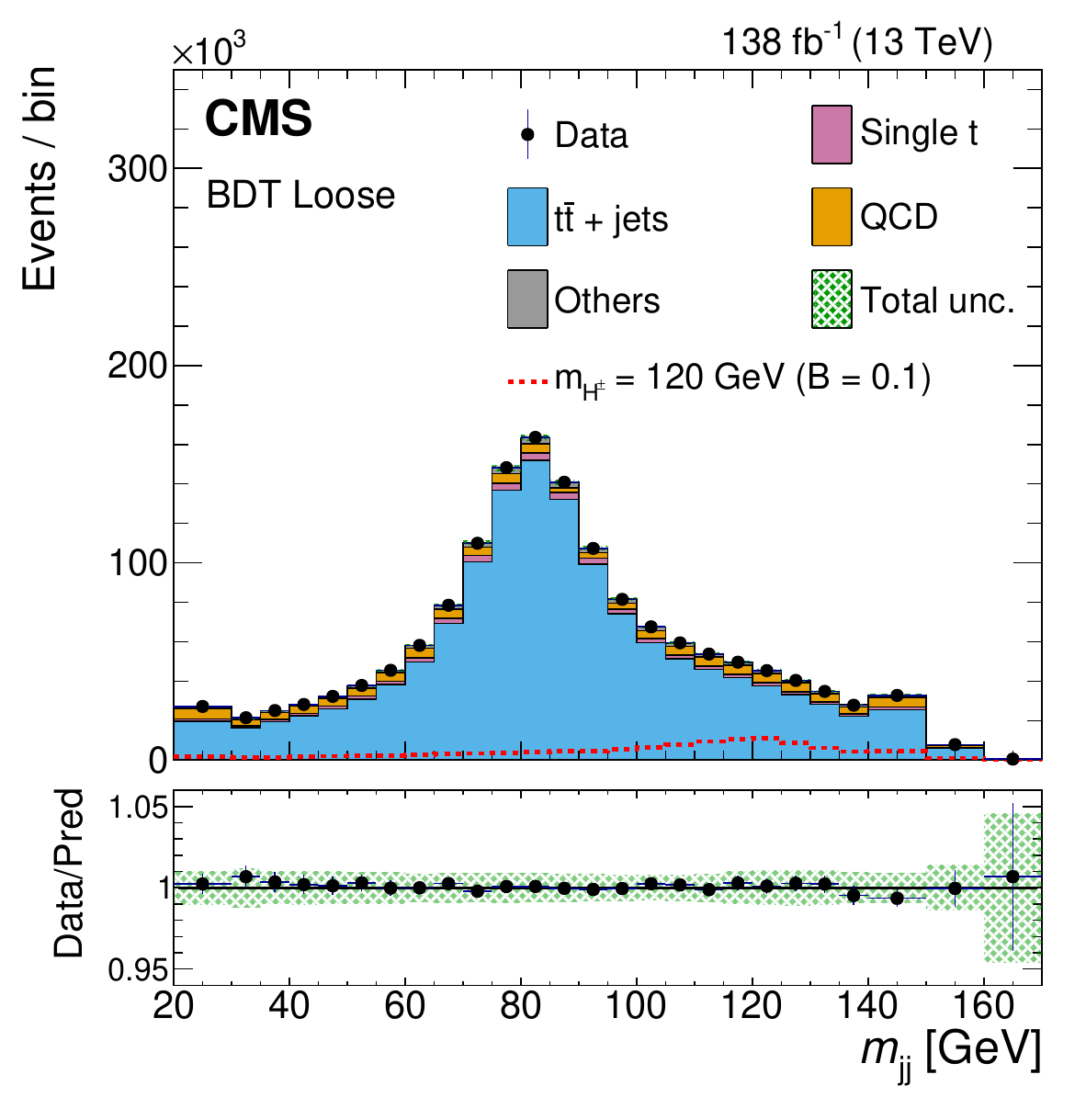}\hfill
  \includegraphics[width=0.48\linewidth]{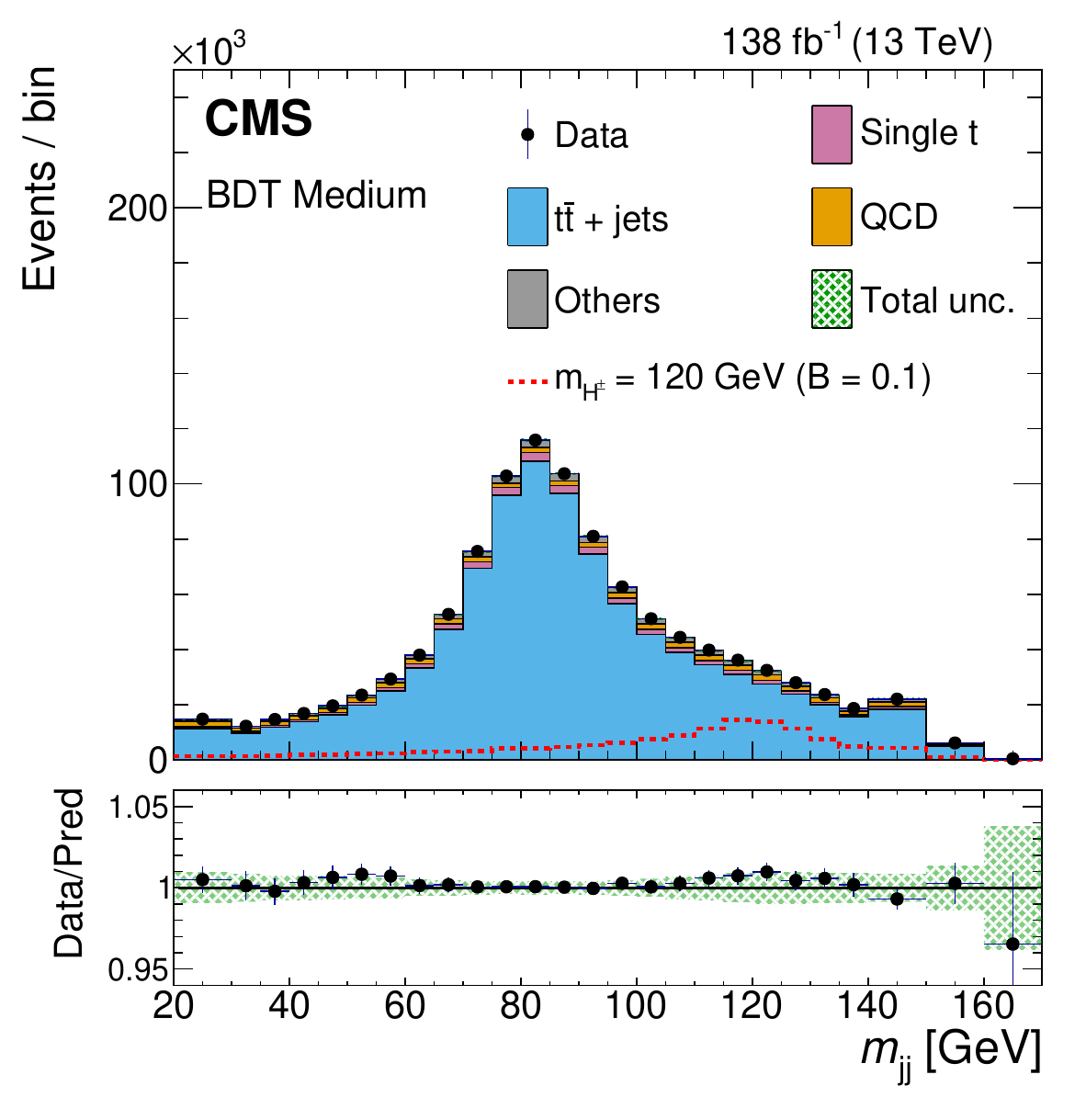}
}
\vspace{1.0ex} 
\begin{center}
  \includegraphics[width=0.48\linewidth]{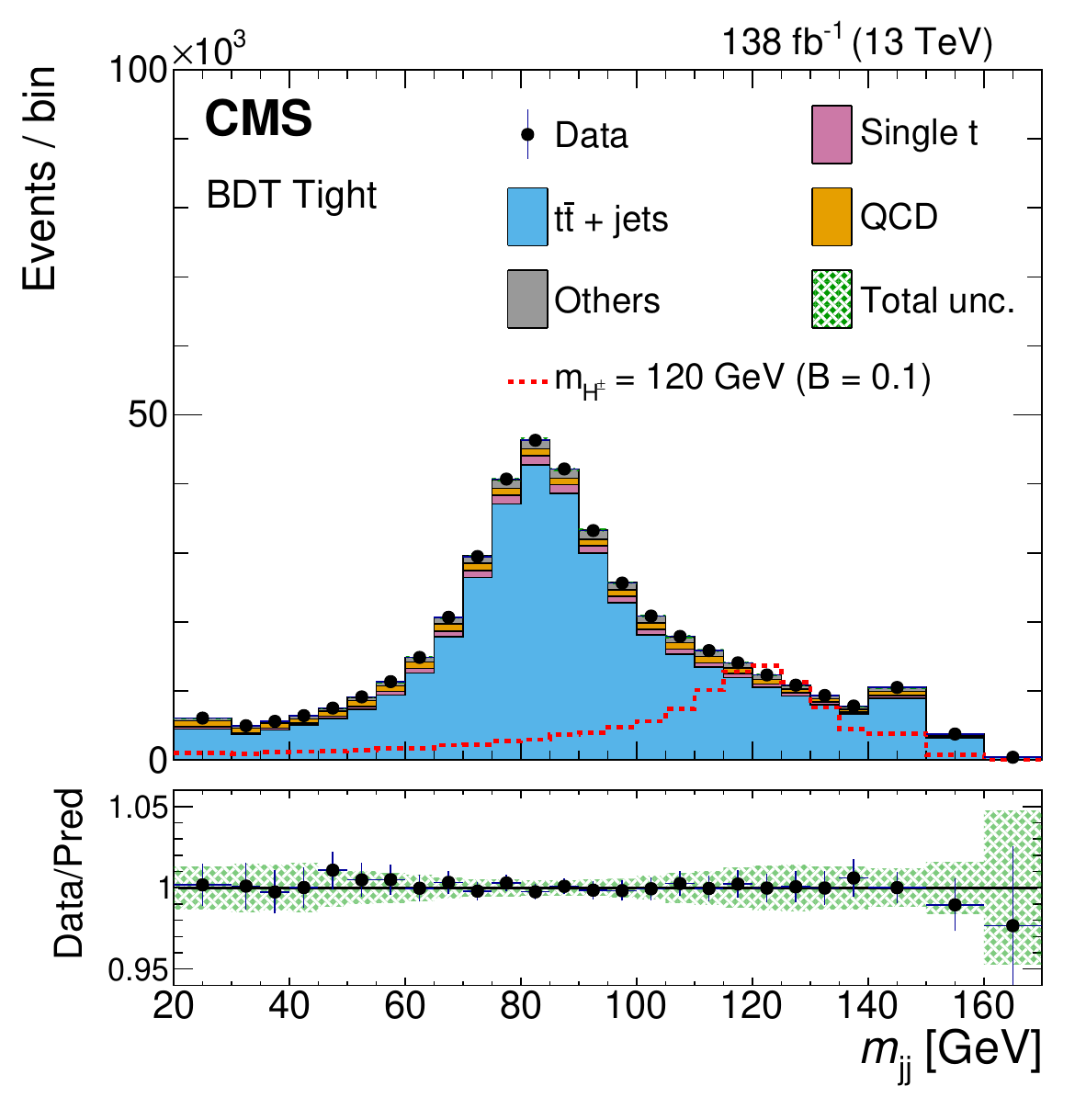}
\end{center}
\caption{The post-fit \mjj\ distributions for data and background processes in loose, medium and tight BDT categories of high-mass model are shown for combined  data set. The \mjj\ variable is defined as the invariant mass of the two light jets associated with the $\PW{}/\PSHpm$ candidate. The potential signal distribution normalized to $\brThb=10\%$  is also overlaid for comparison. The lower panels display the ratio of data to the predicted background, with the shaded band indicating the total uncertainty in the prediction, including both statistical and systematic components. Upper row: loose (left) and medium (right) BDT categories. Bottom row: tight BDT category.}
\label{fig:mjj_cTagEx}
\end{figure*}

Results are presented for both the BDT-based analysis and the alternative cut-based analysis introduced in Section~\ref{ss:secBDT}. The BDT approach provides the best sensitivity, while the cut-based method serves as an independent cross-check. Unless otherwise stated, the results discussed in this section correspond to the BDT-based analysis, with a comparison to the cut-based results provided where relevant. A binned maximum likelihood fit to the \mjj distributions is performed to extract the signal contribution, with background processes constrained within their associated systematic uncertainties. The \mjj distribution is constructed using an adaptive binning scheme to ensure sufficient statistical precision in each bin while maintaining sensitivity to a potential signal resonance. Bins of width 5\GeV are used over most of the mass range, while wider bins are employed in regions with lower event yields: 10\GeV-wide bins in the 140--170\GeV range and in the 20--30\GeV range. The post-fit event yields in data and the SM background estimates are shown in Fig.~\ref{fig:eventYield}. The final \mjj distribution, shown in Fig.~\ref{fig:mjj_cTagEx}, is found to be consistent with the SM background prediction, with the observed resonant structure arising primarily from hadronic \PW boson decays in \ttbar events.

The signal-plus-background model defined below in Eq.~(\ref{nbsm}) accounts for the different \ttbar decay topologies under the hypothesis that the top quark can decay to a charged Higgs boson with \brThb . The total number of produced \ttbar events is fixed, and the possible decay outcomes are treated probabilistically.

Assuming $x$ (representing \brThb) as the parameter of interest, the fit model for signal-plus-background is given by:
\begin{linenomath}
\begin{equation}
\label{nbsm}
N_{ij}(x,\Theta) = x^2 N_{ij}^{\ttbar \to \PSHpm \PSHpm}(\Theta) 
+ 2x(1-x)N_{ij}^{\ttbar \to \PSHpm \PWmp}(\Theta) 
+ (1-x)^2 N_{ij}^{\ttbar\to \PWpm\PWmp}(\Theta) 
+ N_{ij}^{\text{other}}(\Theta),
\end{equation}
\end{linenomath}
and for the background-only case:

\begin{linenomath}
  \begin{equation}
    N_{ij}(\Theta) = N_{ij}^{\ttbar \to \PWpm\PWmp}(\Theta) + N_{ij}^{\text{other}}(\Theta),
  \end{equation}
\end{linenomath}

where $N_{ij}^{\ttbar \to \PSHpm\PSHpm}$, $N_{ij}^{\ttbar \to \PSHpm\PWmp}$, and $N_{ij}^{\ttbar \to \PWpm\PWmp}$ denote the expected event yields in the $i$th bin of the dijet invariant mass distribution for the $j$th charm-tagging or BDT score category, and $\Theta$ represents the set of nuisance parameters associated with systematic and statistical uncertainties affecting the signal and background predictions. A factor of 2 is included in Eq.~(\ref{nbsm}) to account for the charge conjugation of $\PQt$ and $\PAQt$.

Both cases (B-only and S+B) are normalized to the expected \ttbar cross sections.

Although all possible decay configurations are included in this parametrization, the analysis targets lepton+jets final states with a single isolated lepton (electron or muon). As a result, events in which both top quarks decay via $\PQt \to \PSHpm \PQb$ (\ie, events with two charged Higgs bosons)  predominantly lead to fully hadronic final states and therefore fail the event selection. These events do not contribute to the SR. The sole contribution under the signal hypothesis arises from events with one \PSHpm, corresponding to \ttbar events where one top quark decays via $\PQt \to \PSHpm \PQb$ and the other via $\PQt \to \PWpm \PQb$. The background-only hypothesis corresponds to the SM \ttbar contribution along with other background processes.

In this study, the observed events $n_{ij}$ are assumed to follow a Poisson distribution with expected events $N_{ij}(\theta)$. Asymptotic 95\% \CL limits on $\brThb$ are calculated using the $\CLs$ method~\cite{Junk:1999kv,Read:2002hq}, employing a standard likelihood-based approach~\cite{Cowan:2010js}.
The combined maximum-likelihood fits of \mujets and \ejets channels for 2016, 2017, and 2018 data are performed using the CMS statistical analysis tool \textsc{Combine}~\cite{CMS:2024onh}.
The asymptotic limits of the combined results of three years are obtained by the modified frequentist method~\cite{Cowan:2010js,CMS-NOTE-2011-005}.
The profiling of nuisance parameters is performed according to the nature of the associated uncertainty.
Uncertainties related to the integrated luminosity, process normalization, lepton identification
efficiencies, and parton distribution functions are profiled as log-normal constraints, while all remaining systematic uncertainties are treated as shape uncertainties and are profiled accordingly~\cite{Conway:2011in}.

\begin{figure*}[htbp]
        \centering
        \includegraphics[width=0.6\textwidth]{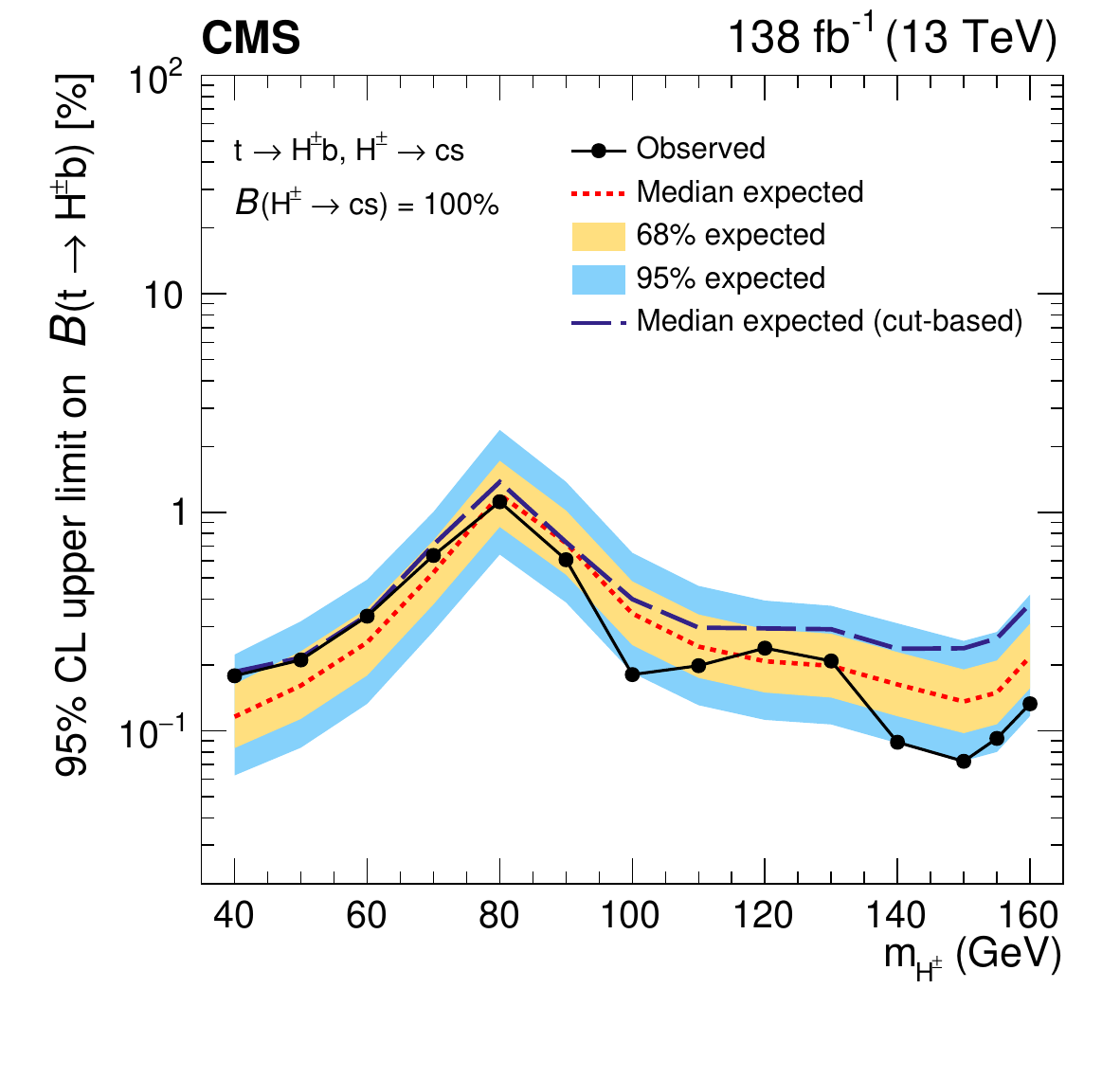}
        \caption{Expected and observed upper limits  at 95\% \CL on the branching fraction \brThb obtained using the BDT-based analysis in the combined lepton category (\mujets and \ejets channels). For comparison, the expected limits from the cut-based analysis are also shown. The connecting lines are for illustration only and do not imply linear interpolation.}
         \label{fig:limitPlot}
\end{figure*}

The exclusion limits are shown in Fig.~\ref{fig:limitPlot}. In all cases, results from the \mujets and \ejets channels are combined. The median expected exclusion limits for the cut-based analysis are also shown. The weakest limits are observed near $m_{\PSHpm} = 80\GeV$, where the \mjj distribution for signal coincides with that of the hadronic \PW boson decays in the dominant \ttbar background. The most stringent limits are obtained around \mhpm = 150\GeV, beyond which the sensitivity degrades due to reduced signal statistics from phase space suppression at higher masses.

The observed limits are up to 75\% better than the previously published CMS results based on 2016 data~\cite{CMS:2020osd}, and approximately 40--70\% better than the corresponding ATLAS results~\cite{ATLAS:2025lightHplus} in the mass range 80--110\GeV, with comparable limits in other masses. While the expected limits are better by 12--48\%  in the range of 70--100\GeV. This improvement is attributed to the larger integrated luminosity, the application of kinematic fitting which improves the \mjj resolution, the enhanced \PQb and \PQc jet tagging performance, and the use of a multivariate (BDT-based) analysis strategy.

\section{Summary}
\label{s:secConcl}

A search has been presented for a light charged Higgs boson (\PSHpm) produced in top quark decays in \ttbar events and decaying to charm and strange quarks using 2016--2018 data recorded by the CMS experiment at the LHC. Upper limits on \brThb are computed at the 95\% confidence level, under the assumption of $\brHcs = 100\%$, with the observed upper limit on \brThb  below 1.1\% for all assumed \PSHpm masses between 40 and 160\GeV. These are the first direct limits on charged Higgs bosons produced in top quark decays for masses between 40 and 50\GeV, and the most stringent limits to date in the 70--110\GeV range.

\begin{acknowledgments}
We congratulate our colleagues in the CERN accelerator departments for the excellent performance of the LHC and thank the technical and administrative staffs at CERN and at other CMS institutes for their contributions to the success of the CMS effort. In addition, we gratefully acknowledge the computing centers and personnel of the Worldwide LHC Computing Grid and other centers for delivering so effectively the computing infrastructure essential to our analyses. Finally, we acknowledge the enduring support for the construction and operation of the LHC, the CMS detector, and the supporting computing infrastructure provided by the following funding agencies: SC (Armenia), BMBWF and FWF (Austria); FNRS and FWO (Belgium); CNPq, CAPES, FAPERJ, FAPERGS, and FAPESP (Brazil); MES and BNSF (Bulgaria); CERN; CAS, MoST, and NSFC (China); MINCIENCIAS (Colombia); MSES and CSF (Croatia); RIF (Cyprus); SENESCYT (Ecuador); ERC PRG and PSG, TARISTU24-TK10 and MoER TK202 (Estonia); Academy of Finland, MEC, and HIP (Finland); CEA and CNRS/IN2P3 (France); SRNSF (Georgia); BMFTR, DFG, and HGF (Germany); GSRI (Greece); MATE and NKFIH (Hungary); DAE and DST (India); IPM (Iran); SFI (Ireland); INFN (Italy); MSIT and NRF (Republic of Korea); MES (Latvia); LMTLT (Lithuania); MOE and UM (Malaysia); BUAP, CINVESTAV, CONACYT, LNS, SEP, and UASLP-FAI (Mexico); MOS (Montenegro); MBIE (New Zealand); PAEC (Pakistan); MSHE, NSC, and NAWA (Poland); FCT (Portugal); MESTD (Serbia); MICIU/AEI and PCTI (Spain); MOSTR (Sri Lanka); Swiss Funding Agencies (Switzerland); MST (Taipei); MHESI (Thailand); TUBITAK and TENMAK (T\"{u}rkiye); NASU (Ukraine); STFC (United Kingdom); DOE and NSF (USA).

\begin{sloppypar}
\setlength\emergencystretch{\hsize}
\hyphenation{Rachada-pisek} Individuals have received support from the Marie-Curie program and the European Research Council and Horizon 2020 Grant, contract Nos.\ 675440, 724704, 752730, 758316, 765710, 824093, 101115353, 101002207, 101001205, and COST Action CA16108 (European Union); the Leventis Foundation; the Alfred P.\ Sloan Foundation; the Alexander von Humboldt Foundation; the Science Committee, project no. 22rl-037 (Armenia); the Fonds pour la Formation \`a la Recherche dans l'Industrie et dans l'Agriculture (FRIA) and Fonds voor Wetenschappelijk Onderzoek contract No. 1228724N (Belgium); the Beijing Municipal Science \& Technology Commission, No. Z191100007219010, the Fundamental Research Funds for the Central Universities, the Ministry of Science and Technology of China under Grant No. 2023YFA1605804, the Natural Science Foundation of China under Grant No. 12535004, and USTC Research Funds of the Double First-Class Initiative No.\ YD2030002017 (China); the Ministry of Education, Youth and Sports (MEYS) of the Czech Republic; the Shota Rustaveli National Science Foundation, grant FR-22-985 (Georgia); the Deutsche Forschungsgemeinschaft (DFG), among others, under Germany's Excellence Strategy -- EXC 2121 ``Quantum Universe" -- 390833306, and under project number 400140256 - GRK2497; the Hellenic Foundation for Research and Innovation (HFRI), Project Number 2288 (Greece); the Hungarian Academy of Sciences, the New National Excellence Program - \'UNKP, the NKFIH research grants K 131991, K 138136, K 143460, K 143477, K 147557, K 146913, K 146914, K 147048, TKP2021-NKTA-64, and 2025-1.1.5-NEMZ\_KI-2025-00004, and MATE KKP and KKPCs Research Excellence and Flagship Research Groups grants (Hungary); the Council of Science and Industrial Research, India; ICSC -- National Research Center for High Performance Computing, Big Data and Quantum Computing, FAIR -- Future Artificial Intelligence Research, and CUP I53D23001070006 (Mission 4 Component 1), funded by the NextGenerationEU program, the Italian Ministry of University and Research (MUR) under Bando PRIN 2022 -- CUP I53C24002390006, PRIN PRIMULA 2022RBYK7T (Italy); the Latvian Council of Science; the Ministry of Science and Higher Education, project no. 2022/WK/14, and the National Science Center, contracts Opus 2021/41/B/ST2/01369, 2021/43/B/ST2/01552, 2023/49/B/ST2/03273, and the NAWA contract BPN/PPO/2021/1/00011 (Poland); the Funda\c{c}\~ao para a Ci\^encia e a Tecnologia (Portugal); the National Priorities Research Program by Qatar National Research Fund; MICIU/AEI/10.13039/501100011033, ERDF/EU, ``European Union NextGenerationEU/PRTR", projects PID2022-142604OB-C21, PID2022-139519OB-C21, PID2023-147706NB-I00, PID2023-148896NB-I00, PID2023-146983NB-I00, PID2023-147115NB-I00, PID2023-148418NB-C41, PID2023-148418NB-C42, PID2023-148418NB-C43, PID2023-148418NB-C44, PID2024-158190NB-C22, RYC2021-033305-I, RYC2024-048719-I, CNS2023-144781, CNS2024-154769 and Plan de Ciencia, Tecnolog{\'i}a e Innovaci{\'o}n de Asturias, Spain; the Chulalongkorn Academic into Its 2nd Century Project Advancement Project, the National Science, Research and Innovation Fund program IND\_FF\_68\_369\_2300\_097, and the Program Management Unit for Human Resources \& Institutional Development, Research and Innovation, grant B39G680009 (Thailand); the Eric \& Wendy Schmidt Fund for Strategic Innovation through the CERN Next Generation Triggers project under grant agreement number SIF-2023-004; the Kavli Foundation; the Nvidia Corporation; the SuperMicro Corporation; the Welch Foundation, contract C-1845; and the Weston Havens Foundation (USA). 
\end{sloppypar} 
\end{acknowledgments}\section*{Data availability} Release and preservation of data used by the CMS Collaboration as the basis for publications is guided by the  \href{https://doi.org/10.7483/OPENDATA.CMS.1BNU.8V1W}{CMS data preservation, re-use and open access policy}.

\bibliography{auto_generated}

\cleardoublepage \appendix\section{The CMS Collaboration \label{app:collab}}\begin{sloppypar}\hyphenpenalty=5000\widowpenalty=500\clubpenalty=5000\cmsinstitute{Yerevan Physics Institute, Yerevan, Armenia}
{\tolerance=6000
A.~Hayrapetyan, V.~Makarenko\cmsorcid{0000-0002-8406-8605}, A.~Tumasyan\cmsAuthorMark{1}\cmsorcid{0009-0000-0684-6742}
\par}
\cmsinstitute{Institut f\"{u}r Hochenergiephysik, Vienna, Austria}
{\tolerance=6000
W.~Adam\cmsorcid{0000-0001-9099-4341}, L.~Benato\cmsorcid{0000-0001-5135-7489}, T.~Bergauer\cmsorcid{0000-0002-5786-0293}, M.~Dragicevic\cmsorcid{0000-0003-1967-6783}, P.S.~Hussain\cmsorcid{0000-0002-4825-5278}, M.~Jeitler\cmsAuthorMark{2}\cmsorcid{0000-0002-5141-9560}, N.~Krammer\cmsorcid{0000-0002-0548-0985}, A.~Li\cmsorcid{0000-0002-4547-116X}, D.~Liko\cmsorcid{0000-0002-3380-473X}, M.~Matthewman, J.~Schieck\cmsAuthorMark{2}\cmsorcid{0000-0002-1058-8093}, R.~Sch\"{o}fbeck\cmsAuthorMark{2}\cmsorcid{0000-0002-2332-8784}, M.~Shooshtari\cmsorcid{0009-0004-8882-4887}, M.~Sonawane\cmsorcid{0000-0003-0510-7010}, N.~Van~Den~Bossche\cmsorcid{0000-0003-2973-4991}, W.~Waltenberger\cmsorcid{0000-0002-6215-7228}, C.E.~Wulz\cmsAuthorMark{2}\cmsorcid{0000-0001-9226-5812}
% ...
\par}
$^{1}$Also at Yerevan State University, Yerevan, Armenia\\
$^{2}$Also at Technische Universit\"{a}t Wien, Vienna, Austria\\
% ...
\end{sloppypar}
\end{document}